\documentclass[aps,amsmath,amssymb,superscriptaddress,prb,preprint]{revtex4-1}

\usepackage[dvipdfmx]{graphicx}
\usepackage{color}

\usepackage{bm}

\usepackage{comment}

\usepackage[normalem]{ulem}

\begin{document}

\title{Generation of orbital currents by magnetization dynamics: Orbital pumping}

\title{Observation of orbital pumping}

\author{Hiroki Hayashi} 
\affiliation{Department of Applied Physics and Physico-Informatics, Keio University, Yokohama 223-8522, Japan}
\affiliation{Keio Institute of Pure and Applied Science, Keio University, Yokohama 223-8522, Japan}

\author{Dongwook Go}
\affiliation{Peter Gr\"unberg Institut and Institute for Advanced Simulation, Forschungszentrum J\"ulich and JARA, 52425 J\"ulich, Germany \looseness=-1}
\affiliation{Institute of Physics, Johannes Gutenberg University Mainz, 55099 Mainz, Germany}

\author{Yuriy Mokrousov}
\affiliation{Peter Gr\"unberg Institut and Institute for Advanced Simulation, Forschungszentrum J\"ulich and JARA, 52425 J\"ulich, Germany \looseness=-1}
\affiliation{Institute of Physics, Johannes Gutenberg University Mainz, 55099 Mainz, Germany}

\author{Kazuya Ando\footnote{Correspondence and requests for materials should be addressed to ando@appi.keio.ac.jp}}
\affiliation{Department of Applied Physics and Physico-Informatics, Keio University, Yokohama 223-8522, Japan}
\affiliation{Keio Institute of Pure and Applied Science, Keio University, Yokohama 223-8522, Japan}
\affiliation{Center for Spintronics Research Network, Keio University, Yokohama 223-8522, Japan}

\maketitle

\bigskip\noindent
\textbf{Abstract\\ 
Harnessing spin and orbital angular momentum is a fundamental concept in condensed matter physics, materials science, and quantum-device applications. In particular, the search for phenomena that generate a flow of spin angular momentum, a spin current, has led to the development of spintronics, advancing the understanding of angular momentum dynamics at the nanoscale. In contrast, the generation of an orbital current, the orbital counterpart of a spin current, remains a significant challenge. Here, we report the observation of orbital-current generation from magnetization dynamics: orbital pumping. We show that the orbital pumping in Ni/Ti bilayers injects an orbital current into the Ti layer, which is detected through the inverse orbital Hall effect. The orbital pumping is the orbital counterpart of the spin pumping, which is one of the most versatile and powerful mechanisms for spin-current generation. Our findings provide a promising approach for generating orbital currents, opening the door to explore the orbital analogue of spintronics: orbitronics.
}

\bigskip\noindent
\textbf{Main}\\
Electrons carry spin and orbital angular momentum, 
both of which are fundamental to the electronic and magnetic properties of materials. 
Over the past three decades, studies of a flow of spin angular momentum, a spin current, have revealed its fascinating topological, relativistic, and quantum mechanical nature, leading to the development of spintronics. 
The key element of spintronics is the interplay between spin currents and magnetization in ferromagnetic/nonmagnetic (FM/NM) bilayers~\cite{oxfordspin}. 
In a FM/NM bilayer, when a spin current is injected into the FM layer from the adjacent NM layer, the injected spin current interacts with the magnetization through the exchange coupling in the FM layer, exerting a torque, called a spin torque, on the magnetization~\cite{RevModPhys.91.035004}. The spin torque drives domain-wall motion, magnetization switching, and magnetization precession, providing a way to realize a plethora of ultralow power and fast spintronic devices, such as nonvolatile magnetic memories, nanoscale microwave sources, and neuromorphic computing devices~\cite{ryu2020current,hirohata2020review,dieny2020opportunities}. 
The reciprocal effect of the spin torque is known as the spin pumping, a phenomenon in which precessing magnetization in the FM layer pumps a spin current into the adjacent NM layer~\cite{Mizukami,Tserkovnyak1,Tserkovnyak_rev}.
The spin pumping is one of the most versatile and powerful mechanisms for spin-current generation, leading to the discovery of a variety of spin-current phenomena, such as spin transport in insulators and nonvolatile electric control of spin-charge conversion~\cite{kajiwara,PhysRevLett.113.097202,noel2020non}. 
This pair of reciprocal effects, the spin torque and the spin pumping, has played a central role in the development of modern spintronics.

\begin{figure*}[tb]
\center\includegraphics[scale=1.0]{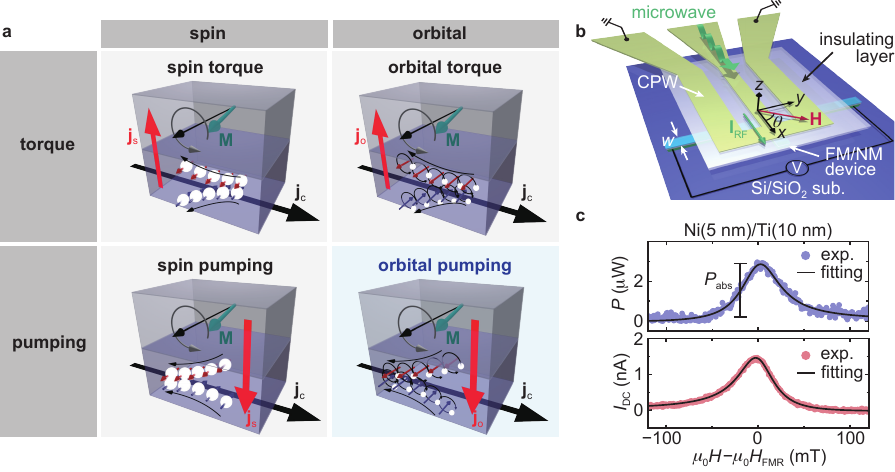}
\caption{
{\bfseries Orbital pumping and inverse orbital Hall effect.} 
\textbf{a}, Schematic illustration of the spin torque, spin pumping, orbital torque, and orbital pumping. 
\textbf{b}, Schematic illustration of the experimental setup. 
\textbf{c}, Magnetic field $H$ dependence of the microwave absorption intensity $P$ (upper) and the charge current $I_\mathrm{DC}$ (lower) for the Ni(5~nm)/Ti(10~nm) bilayer measured at $\theta=0$ with $f=10$~GHz and $P_\mathrm{in}=10$~mW. The solid circles are the experimental data. The $H$ dependence of $I_\mathrm{DC}$ is fitted with the sum of symmetric and antisymmetric functions (see the solid curve). The $H$ dependence of $P$ is fitted with the sum of symmetric and antisymmetric functions~\cite{doi:10.1063/1.5007943} (see the solid curve): $P=P_\text{abs}{W^2}/\left[{(\mu_0 H-\mu_0 H_\text{FMR})^2+W^2}\right]+P_\text{antisym}W(\mu_0 H-\mu_0 H_\text{FMR})/\left[{(\mu_0 H-\mu_0 H_\text{FMR})^2+W^2}\right]$.}
\label{fig1}
\end{figure*}

In contrast to the success in establishing the physics of spin currents, the orbital counterpart of spin currents, orbital currents, has been largely overlooked. Recent studies, however, have suggested the crucial role of orbital currents, a flow of orbital angular momentum, in the dynamics of angular momentum in solids~\cite{go2021orbitronics,KIM2022169974}. 
In particular, experimental and theoretical studies have suggested the existence of the orbital counterpart of the spin torque: an orbital torque~\cite{PhysRevResearch.2.013177,PtCo-orbital,PhysRevB.103.L020407,tazaki2020current,PhysRevLett.125.177201,Cr-orbital,Ta-orbital,choi2021observation,hayashi2022observation,PhysRevResearch.4.033037}. The orbital torque emerges when an orbital current is injected into a FM layer. In the FM layer, the injected orbital angular momentum interacts with the magnetization through a combined action of the spin-orbit coupling and the exchange coupling in the FM layer, exerting a torque on the magnetization, which is the orbital torque~\cite{PhysRevResearch.2.013177}. 
The Onsager's reciprocal relations guarantee the existence of the reciprocal effect of the orbital torque, in which an orbital current is pumped by magnetization dynamics. This phenomenon is the orbital counterpart of the spin pumping and can be referred to as orbital pumping. 
Despite its importance, however, the orbital pumping remains unexplored; the orbital pumping is the missing piece in the set of spin and orbital current phenomena (see Fig.~\ref{fig1}a).

In this work, we report the observation of the orbital pumping. We show that a charge current is generated at ferromagnetic resonance (FMR) in Ni/Ti bilayers, in which the existence of the orbital torque originating from the orbital Hall effect (OHE) is well established~\cite{hayashi2022observation}. In FM/NM bilayers, the precession of the magnetization in the FM layer can generate charge currents in the NM layer through two distinct channels: the spin pumping followed by the inverse spin Hall effect (ISHE) and the orbital pumping followed by the inverse orbital Hall effect (IOHE). We demonstrate that the experimental characteristics of the observed current are consistent with it being generated through the IOHE induced by the orbital pumping.

The orbital pumping can be understood as an orbital analog of the spin pumping~\cite{Tserkovnyak1} with the same symmetry constraint (the physical picture is explained in Supplementary Note 1). In a NM/FM bilayer, an orbital current driven by dynamics of the magnetization is given by 
\begin{equation}
{\bf j}_\mathrm{o}=\beta {\bf m}\times \frac{d{\bf m}}{dt} + \beta' \frac{d{\bf m}}{dt}, 
\label{eq:OP}
\end{equation}
where ${\bf m}$ denotes the unit vector of the magnetization. Here, the orbital polarization direction of ${\bf j}_\mathrm{o}$ is represented by its vector direction, and the propagation direction of electrons is perpendicular to the interface. However, the crucial difference between the orbital pumping and the spin pumping is that the orbital pumping requires the spin-orbit coupling in its microscopic mechanism~\cite{arXiv:2309.14817}. This is because the orbital angular momentum interacts with the magnetization only indirectly via the spin-orbit interaction~\cite{PhysRevResearch.2.013177, PhysRevResearch.2.033401}. Therefore, the orbital pumping sensitively depends on how the spin and orbital degrees of freedom are correlated, which differs from material to material~\cite{arXiv:2309.14817}. A recent theory predicts that ferromagnetic Ni exhibits the strongest orbital pumping among Fe, Co, and Ni~\cite{arXiv:2309.14817}. It has been found that Ni exhibits pronounced correlation between the spin and orbital angular momenta near the Fermi surface~\cite{Ta-orbital, PhysRevResearch.2.033401,arXiv:2309.14817}.

As schematically illustrated in Fig.~\ref{fig1}a, the orbital pumping can be electrically detected by the IOHE, as the spin pumping is detected by the ISHE. Here, the major difference between the IOHE and ISHE is their dependence on the spin-orbit coupling. While the ISHE arises from the spin-orbit coupling, the IOHE does not require it. Thus, a light metal such as Ti can be used to detect the orbital pumping. We note that recent experiments have revealed large OHE in Ti~\cite{hayashi2022observation, choi2021observation}. To detect the spin pumping, on the other hand, typically a heavy metal such as Pt is employed for its large ISHE. This motivates us to consider the Ti/Ni bilayers, in which gigantic orbital torque response has been observed in the previous work~\cite{hayashi2022observation}.

\bigskip\noindent
\textbf{Charge-current generation}\\
To detect the orbital pumping, we measured the direct-current (DC) voltage $V_\mathrm{DC}$ generated at the FMR for FM/NM bilayers with FM = Ni, Fe, or Co and NM = Ti or Pt. In Fig.~\ref{fig1}b, we show a schematic of the device and experimental setup (for details, see Methods). 
Equation~(\ref{eq:OP}) shows that the DC component of an orbital current generated by the orbital pumping is proportional to the projection of ${\bf m}\times d{\bf m}/dt$ onto the magnetization-precession axis. This projection is proportional to the square of the magnetization-precession amplitude, showing that the injected orbital current is proportional to the microwave power $P_\mathrm{in}$ and microwave absorption intensity $P$ (see also Supplementary Note 2). 
Equation~(\ref{eq:OP}) also shows that the orbital polarization direction of the DC component of the injected orbital current is directed along the magnetization-precession axis, which is parallel to an external magnetic field in the FM/NM bilayers at the FMR. 
For the measurement, we applied a radio frequency (RF) current $I_\mathrm{RF}$ with a frequency of $f$ perpendicular to the direction across electrodes attached to the FM/NM device and swept an in-plane external field $H$ applied at an angle of $\theta$ from the direction of the applied RF current (see Fig.~\ref{fig1}b).

In Fig.~\ref{fig1}c, we show magnetic field $H$ dependence of the microwave absorption $P$ and charge current $I_\mathrm{DC}=V_\mathrm{DC}/R$ signals measured for the Ni(5~nm)/Ti(10~nm) bilayer at $\theta=0$, where $R$ is the resistance of the device between the electrodes, and the numbers in parentheses represent the thickness.  
Figure~\ref{fig1}c shows that despite the negligible ISHE in the Ti layer~\cite{du2014systematic}, a clear $I_\mathrm{DC}$ signal appears around the FMR field $H_\mathrm{FMR}$. The observed charge-current signal can be decomposed into symmetric and antisymmetric functions with respect to $H_\mathrm{FMR}$: $I_\mathrm{DC}=I_\text{sym}{W^2}/\left[{(\mu_0 H-\mu_0 H_\text{FMR})^2+W^2}\right]+I_\text{antisym}W(\mu_0 H-\mu_0 H_\text{FMR})/\left[{(\mu_0 H-\mu_0 H_\text{FMR})^2+W^2}\right]$, where $W$ denotes the FMR linewidth~\cite{Saitoh}. 
The spin and orbital pumping generate the symmetric component $I_\text{sym}$ through the ISHE and IOHE, respectively, because the DC component of the pumped current is proportional to the microwave absorption intensity~\cite{ando2011inverse}. 
Figure~\ref{fig1}c shows that the $I_\mathrm{DC}$ signal in the Ni/Ti bilayer is dominated by the symmetric component $I_\mathrm{sym}$, suggesting that the observed charge current stems from the orbital pumping. 
Here, in the $I_\mathrm{DC}$ signal, the nonzero antisymmetric component $I_\text{antisym}$ arises from spin rectification effects, which can be induced by the anisotropic magnetoresistance (AMR), the planar Hall effect (PHE), and the anomalous Hall effect (AHE) in the Ni layer~\cite{PhysRevB.84.054423,doi:10.7566/JPSJ.86.011003}. The asymmetric line shape in the $P$ signal is presumably induced by eddy currents, which lead to an absorption-dispersion admixture~\cite{10.1063/1.4917285}.

\begin{figure*}[tb]
\center\includegraphics[scale=1.0]{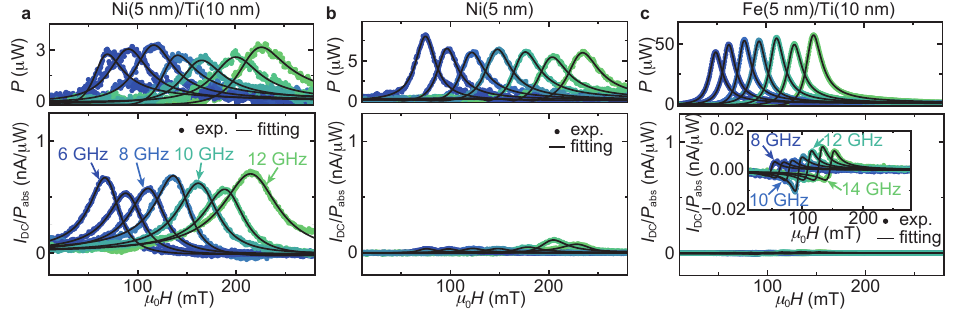}
\caption{
{\bfseries Role of Ti and FM layers in charge-current generation.}
\textbf{a}-\textbf{c}, Magnetic field $H$ dependence of $P$ and $I_\mathrm{DC}/P_\mathrm{abs}$ for (\textbf{a}) the Ni(5~nm)/Ti(10~nm) bilayer, (\textbf{b}) the Ni(5~nm) single-layer, and (\textbf{c}) the Fe(5~nm)/Ti(10~nm) bilayer measured at $\theta=0$ and $P_\mathrm{in}=10$~mW. The frequency $f$ of the RF current was varied from 6~GHz to 12~GHz in 1~GHz steps for the Ni/Ti and Ni films and from 8~GHz to 14~GHz in 1~GHz steps for the Fe/Ti film. The charge current is defined as $I_\mathrm{DC}=V_\mathrm{DC}/R$, and $P_\mathrm{abs}$ is the microwave absorption intensity at $H=H_\mathrm{FMR}$ (see Fig.~\ref{fig1}c). The solid circles are the experimental data, and the solid curves are the fitting result using the sum of symmetric and antisymmetric functions. The inset shows a magnified view of the signals.}
\label{fig2}
\end{figure*}

To reveal the origin of the charge current observed in the Ni/Ti bilayer, we measured the microwave absorption $P$ and the charge current $I_\mathrm{DC}$ for Ni(5~nm) and Fe(5~nm)/Ti(10~nm) films, as well as the Ni(5~nm)/Ti(10~nm) bilayer at various frequencies $f$. Figures~\ref{fig2}a-\ref{fig2}c show the $H$ dependence of $P$ and $I_\mathrm{DC}/P_\mathrm{abs}$ for the three devices at $\theta=0$, where $P_\mathrm{abs}$ is the microwave absorption intensity at the FMR field (see Fig.~\ref{fig1}c). 
We note that the symmetric component of $I_\mathrm{DC}/P_\mathrm{abs}$, $I_\mathrm{sym}/P_\mathrm{abs}$, in the Ni single-layer film is negligible compared to that in the Ni/Ti bilayer (see Figs.~\ref{fig2}a and \ref{fig2}b). This result demonstrates that the $I_\mathrm{sym}$ signal in the Ni/Ti bilayer is generated in the Ti layer; spin-rectification effects and thermoelectric effects in the Ni layer are not the source of the observed $I_\mathrm{sym}$ signals in the Ni/Ti bilayer. The contribution from the charge current generated in the Ni layer is discussed in detail in Supplementary Note 3.

An important feature of the $I_\mathrm{sym}$ signal is that its magnitude strongly depends on the choice of the FM layer. 
Figure~\ref{fig2}c shows that $I_\mathrm{sym}/P_\mathrm{abs}$ is vanishingly small in the Fe/Ti bilayer, despite the fact that the $I_\mathrm{sym}$ signal in the Ni/Ti bilayer originates from the Ti layer. The distinct difference in the magnitude of $I_\mathrm{sym}/P_\mathrm{abs}$ between the Ni/Ti and Fe/Ti bilayers is the key feature of the orbital pumping~\cite{arXiv:2309.14817}. 
In Ni, the electronic occupation of the $d$ orbital shells is optimized such that the spin-orbit correlation is particularly strong near the Fermi energy~\cite{PhysRevResearch.2.033401,Ta-orbital,hayashi2022observation}. The strong spin-orbit correlation near the Fermi energy results in the strong coupling between the magnetization and orbital current in Ni, which has been confirmed by the orbital-torque studies~\cite{Ta-orbital,Cr-orbital,hayashi2022observation,PhysRevResearch.4.033037}. This implies a large orbital pumping efficiency $\beta$ in the Ni/Ti bilayer (see Eq.~(\ref{eq:OP})), suggesting that the orbital pumping in the Ni/Ti bilayer can inject a large orbital current into the Ti bilayer. 
In contrast, we expect that the orbital pumping in the Fe/Ti bilayer is inefficient. The reason for this is that the spin-orbit correlation near the Fermi energy in Fe is much smaller than that in Ni, as the hotspots for the spin-orbit correlation are located about 1~eV below the Fermi energy in Fe~\cite{PhysRevResearch.2.033401}. The inefficient coupling between the magnetization and orbital current in Fe has been demonstrated in previous reports; while the current-induced torques in Ni-based structures are dominated by the orbital torques, the spin torques provide the dominant contribution in Fe-based structures~\cite{Ta-orbital,hayashi2022observation}. 
This indicates that the distinct difference in the observed charge-current signals between the Ni/Ti and Fe/Ti bilayers can be primarily attributed to the large difference in the orbital pumping efficiency $\beta$ due to the different strengths of the spin-orbit correlation near the Fermi energy between Ni and Fe (see also Supplementary Note 1).

We have confirmed that the $f$ dependence of $I_\mathrm{sym}/P_\mathrm{abs}$ of the Ni/Ti bilayer is consistent with the model of the spin and orbital pumping (see Supplementary Note 4).
Here, the negligible $I_\mathrm{sym}/P_\mathrm{abs}$ in the Fe/Ti bilayer indicates that the charge current due to the ISHE is negligible in the FM/Ti bilayers. This result is consistent with the vanishingly small spin Hall angle of Ti~\cite{du2014systematic}, $\theta_\mathrm{SH}=-3.6\times 10^{-4}$, which is three orders of magnitude smaller than that of Pt.
The negligible $I_\mathrm{sym}/P_\mathrm{abs}$ in the Fe/Ti bilayer also supports that sample heating is irrelevant to the charge-current signals observed in the Ni/Ti bilayer because thermoelectric effects are comparable between Fe and Ni~\cite{PhysRevB.96.174406}.

\begin{figure}[tb]
\center\includegraphics[scale=1.0]{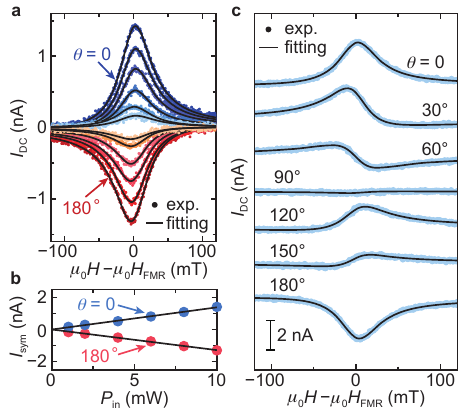}
\caption{
{\bfseries Charge-current spectra at various microwave powers and magnetic field angles.} 
\textbf{a}, Magnetic field $H$ dependence of $I_\mathrm{DC}$ for the Ni(5~nm)/Ti(10~nm) bilayer at $\theta =0$ (blue) and $\theta=180^\circ$ (red) measured with $f=10$~GHz and different microwave powers $P_\mathrm{in}=1$, 2, 4, 6, 8 and 10~mW. The solid circles are the experimental data, and the solid curves are the fitting result using the sum of symmetric and antisymmetric functions.
\textbf{b}, $P_\mathrm{in}$ dependence of $I_\mathrm{sym}$ for the Ni(5~nm)/Ti(10~nm) bilayer measured with $f=10$~GHz at $\theta =0$ (blue) and $\theta=180^\circ$ (red). The solid circles are the experimental data, and the solid lines are the linear fitting result.
\textbf{c}, $H$ dependence of $I_\mathrm{DC}$ for the Ni(5~nm)/Ti(10~nm) bilayer at different in-plane magnetic field angle $\theta$ measured with $P_\mathrm{in}=10$~mW and $f=10$~GHz. 
}
\label{fig3}
\end{figure}

\bigskip\noindent
\textbf{Characterization of charge current}\\
To obtain further evidence for the IOHE induced by the orbital pumping, we measured $I_\mathrm{DC}$ for the Ni/Ti bilayer by varying the microwave power $P_\mathrm{in}$ and magnetic field angle $\theta$. 
In Fig.~\ref{fig3}a, we show the $I_\mathrm{DC}$ spectra for the Ni/Ti bilayer measured at different $P_\mathrm{in}$ with applying the magnetic field at $\theta=0$ and $\theta=180^\circ$. By fitting the spectra using the sum of the symmetric and antisymmetric functions, we obtain the $P_\mathrm{in}$ dependence of $I_\mathrm{sym}$, as shown in Fig.~\ref{fig3}b. This result indicates that $I_\mathrm{sym}$ is proportional to $P_\mathrm{in}$, which is consistent with the scenario of the DC orbital pumping.

\begin{figure*}[tb]
\center\includegraphics[scale=1.0]{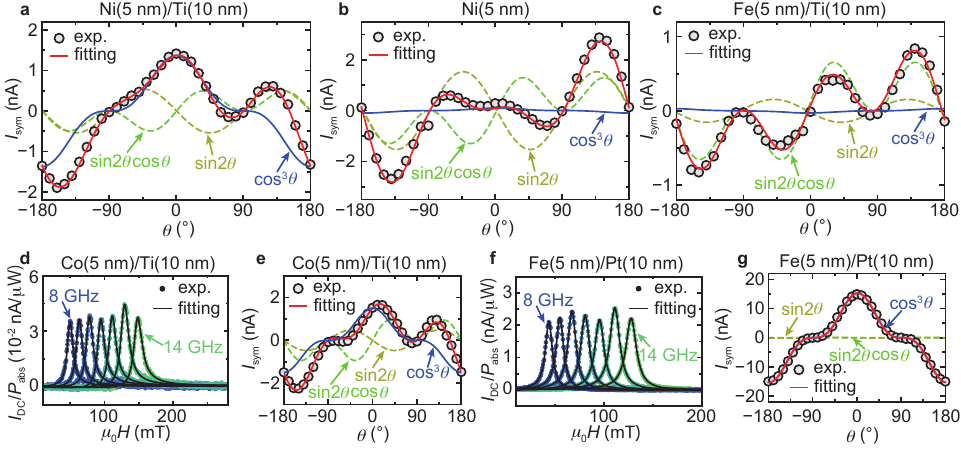}
\caption{
{\bfseries Magnetic-field angle dependence of charge current.}
\textbf{a}-\textbf{c}, In-plane magnetic field angle $\theta$ dependence of $I_\mathrm{sym}$ for (\textbf{a}) the Ni(5~nm)/Ti(10~nm) bilayer, (\textbf{b}) the Ni(5~nm) single-layer, and (\textbf{c}) the Fe(5~nm)/Ti(10~nm) bilayer at $f= 10$~GHz with $P_\mathrm{in}=10$~mW. The curves in red are the fitting result using Eq.~(\ref{eq:angle}), which is the sum of functions proportional to $\cos^3\theta$ (solid curves in blue), $\sin2\theta\cos\theta$ (dashed curves in green), and  $\sin2\theta$ (dashed curves in yellow). 
\textbf{d}, $H$ dependence of $I_\mathrm{DC}/P_\mathrm{abs}$ for the Co(5~nm)/Ti(10~nm) bilayer measured at $\theta=0$ and $P_\mathrm{in}=10$~mW. The frequency $f$ of the RF current was varied from 8~GHz to 14~GHz in 1~GHz steps.
\textbf{e}, $\theta$ dependence of $I_\mathrm{sym}$ for the Co(5~nm)/Ti(10~nm) bilayer at $f= 10$~GHz. 
\textbf{f}, $H$ dependence of $I_\mathrm{DC}/P_\mathrm{abs}$ for the Fe(5~nm)/Pt(10~nm) bilayer measured at $\theta=0$ and $P_\mathrm{in}=1$~mW. The frequency $f$ of the RF current was varied from 8~GHz to 14~GHz in 1~GHz steps.
\textbf{g}, $\theta$ dependence of $I_\mathrm{sym}$ for the Fe(5~nm)/Pt(10~nm) bilayer at $f= 10$~GHz. 
}
\label{fig4}
\end{figure*}

Figures~\ref{fig3}a and \ref{fig3}b demonstrate that the sign of $I_\mathrm{sym}$ is reversed by reversing the magnetic field direction. This result is also consistent with the prediction of the orbital pumping because the DC component of the orbital polarization direction is reversed by reversing the magnetic field direction. We further studied the magnetic field angle $\theta$ dependence of the charge-current generation, as shown in Fig.~\ref{fig3}c. 
By fitting the measured spectra, we plot $I_\mathrm{sym}$ with respect to $\theta$ for the Ni/Ti bilayer in Fig.~\ref{fig4}a. 
As shown in Fig.~\ref{fig4}a, the $\theta$ dependence of $I_\mathrm{sym}$ is well fitted by~\cite{PhysRevB.84.054423,PhysRevB.94.134421,HARDER20161,doi:10.7566/JPSJ.86.011003}
\begin{equation}
I_\mathrm{sym}=I_\mathrm{pump} \cos^3\theta + I_\mathrm{AMR}^\parallel \sin2\theta\cos\theta+I_\mathrm{AMR}^\perp\sin2\theta, \label{eq:angle}
\end{equation}
where $I_\mathrm{pump}$ arises from the spin or orbital pumping. $I_\mathrm{AMR}^\parallel$ and $I_\mathrm{AMR}^\perp$ arise from the AMR due to in-plane and out-of-plane microwave magnetic fields, respectively. 
This result shows that the spin rectification signal in this system is dominated by the AMR because the PHE and AHE contributions show different angular dependences; the PHE contributions are proportional to $\cos 2\theta\cos\theta$ and $\cos 2\theta$ for in-plane and out-of-plane microwave magnetic fields, respectively, and the AHE contribution varies as $\cos \theta$ for an in-plane microwave magnetic field and is constant for an out-of-plane microwave magnetic field~\cite{HARDER20161,doi:10.7566/JPSJ.86.011003}.
Here, Eq.~(\ref{eq:angle}) indicates that $I_\mathrm{sym}$ at $\theta=0$ corresponds to $I_\mathrm{pump}$. 
The fitting result in Fig.~\ref{fig4}a confirms that the $I_\mathrm{sym}$ component of the charge-current signal in the Ni/Ti bilayer at $\theta=0$ varies as $\cos^3\theta$, supporting that the charge-current signals shown in Fig.~\ref{fig2}a arise from the orbital pumping. We also confirmed that the $I_\mathrm{pump}$ component is negligible in the Ni single-layer and the Fe/Ti bilayer, as shown in Figs.~\ref{fig4}b and \ref{fig4}c. These results support that the generation of the charge current requires the IOHE of the Ti layer and the strong spin-orbit correlation of the Ni layer. 
Here, we note that $I_\mathrm{pump}$ is also observed in a Co/Ti bilayer (see Figs.~\ref{fig4}d and \ref{fig4}e). The measured value of $I_\mathrm{pump}/P_\mathrm{abs}$ for the Co/Ti bilayer is smaller than that of the Ni/Ti bilayer but is larger than that of the Fe/Ti bilayer. This result is consistent with the model of the orbital pumping because the orbital response of Co is weaker than that of Ni but is stronger than that of Fe~\cite{Ta-orbital}.
For the Ni/Ti bilayer, we have further checked the validity of the experiment by performing the pumping experiment in which the out-of-plane microwave magnetic field is dominant (see Supplementary Note 5). The magnetic damping is also investigated for the Ni/Ti bilayer (see Supplementary Note 6).

As a reference, we measured $I_\mathrm{DC}$ for a Fe(5~nm)/Pt(10~nm) bilayer, as shown in Figs.~\ref{fig4}f and \ref{fig4}g. This result shows that the sign of $I_\mathrm{pump}$ in the Ni/Ti bilayer is the same as that in the Fe/Pt bilayer. 
In the Fe/Pt bilayer, the $I_\mathrm{pump}$ signal is dominated by the ISHE due to the spin pumping because of the weak spin-orbit correlation in the Fe layer and the strong ISHE in the Pt layer. 
In the scenario of the spin pumping, the spin polarization direction is defined by the magnetization direction and is independent of the choice of the FM layer~\cite{yoshino2011universality}. In contrast, the orbital polarization direction of the orbital current generated by the orbital pumping can depend on the sign of the spin-orbit correlation in the FM layer. In Ni, the spin-orbit correlation is positive, indicating that the direction of the orbital polarization of the orbital current is the same as that of the spin polarization of the spin current generated by the spin pumping~\cite{PhysRevResearch.2.013177}. Thus, the same sign of $I_\mathrm{pump}$ between the Ni/Ti and Fe/Pt bilayers indicates that the sign of the orbital Hall angle in the Ti layer is the same as that of the spin Hall angle in the Pt layer. This result is consistent with the theoretical prediction that the signs of the orbital Hall conductivity in Ti and the spin Hall conductivity in Pt are both positive~\cite{PhysRevMaterials.6.095001}, supporting that the observed signal in the Ni/Ti bilayer arises from the IOHE. 
We also note that the sign of the orbital torque in the Ni/Ti bilayer is the same as that of the spin torque produced by the SHE in Pt~\cite{hayashi2022observation}.

To characterize the strength of the charge-current generation through the orbital pumping and IOHE, we define the conversion efficiency from the absorbed microwave power $P_\mathrm{abs}$ to the generated charge current $I_\mathrm{pump}$ as 
\begin{equation}
\kappa = \frac{ I_\mathrm{pump}}{P_\mathrm{abs}}   
\frac{\xi\gamma W M_\mathrm{s}t_\mathrm{FM}\sigma_\mathrm{NM}\sqrt{(\mu_0 M_\mathrm{eff}^2)+(4\pi f/\gamma)^2}}{2e f}, \label{eq:eff}
\end{equation}
where $\gamma$ is the gyromagnetic ratio, $e$ is the elementary charge, and $\xi$ is a geometrical factor (for details, see Supplementary Note 2). 
$t_\mathrm{FM}$, $M_\mathrm{s}$, and $M_\mathrm{eff}$ are the thickness, saturation magnetization, and effective demagnetization field of the FM layer, respectively. 
$\sigma_\mathrm{NM}$ is the conductivity of the NM layer. Here, the conversion efficiency $\kappa$ is defined such that it corresponds to $g_{\rm{eff}}^{\uparrow\downarrow}\sigma_{\mathrm{SHE}}    \lambda_{\mathrm{s}}\tanh \left({t_{\mathrm{NM}}}/{2 \lambda_{\mathrm{s}}}\right)$ in the case where the charge current originates from the ISHE induced by the spin pumping, where $g_{\rm{eff}}^{\uparrow\downarrow}$ is the effective spin mixing conductance, $\sigma_\mathrm{SHE}$ is the spin Hall conductivity, and $\lambda_\mathrm{s}$ is the spin diffusion length. 
In analogy to the spin pumping, we expect that the charge current generated by the orbital pumping is determined by the orbital pumping efficiency $\beta$, the orbital Hall conductivity $\sigma_\mathrm{OHE}$, and the orbital diffusion length $\lambda_\mathrm{o}$, and therefore we use $\kappa$ defined in Eq.~(\ref{eq:eff}) to characterize the strength of the orbital pumping and IOHE.

\begin{figure}[tb]
\center\includegraphics[scale=1.0]{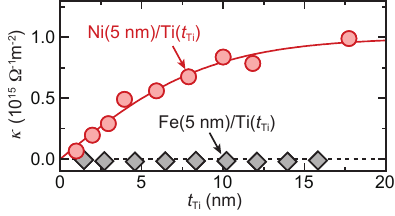}
\caption{
{\bfseries Charge-current generation efficiency.} Ti-layer thickness $t_\mathrm{Ti}$ dependence of the charge-current generation efficiency $\kappa$, defined in Eq.~(\ref{eq:eff}), for the Ni(5~nm)/Ti($t_\mathrm{Ti}$) bilayers (red circles) and the Fe(5~nm)/Ti($t_\mathrm{Ti}$) bilayers (gray diamonds). The values of $\kappa$ are determined from the $I_\mathrm{pump}$ and $P_\mathrm{abs}$ measured at $f=10$~GHz, where $I_\mathrm{pump}$ is determined from the $\theta$ dependence of $I_\mathrm{DC}$ for each device. The solid curve in red is the fitting result using a function proportional to $\tanh(t_\mathrm{Ti}/2\lambda_\mathrm{Ti})$.
}
\label{fig5}
\end{figure}

In Fig.~\ref{fig5}, we show Ti-layer thickness $t_\mathrm{Ti}$ dependence of $\kappa$, determined by measuring the $\theta$ dependence of $I_\mathrm{DC}$, for the Ni(5~nm)/Ti($t_\mathrm{Ti}$) bilayers. 
The charge-current generation efficiency $\kappa$ of the Ni/Ti bilayer is about an order of magnitude smaller than $\kappa=24\times 10^{15}$~$\Omega^{-1}$m$^{-2}$ of the Fe(5~nm)/Pt(10~nm) bilayer, which is roughly consistent with the difference between the generation efficiencies of the orbital and spin torques in these systems~\cite{hayashi2022observation}.
Figure~\ref{fig5} shows that $\kappa$ of the Ni/Ti bilayer increases with $t_\mathrm{Ti}$. 
By fitting the $t_\mathrm{Ti}$ dependence of $\kappa$ for the Ni/Ti bilayer with a function proportional to $\tanh(t_\mathrm{Ti}/2\lambda_\mathrm{Ti})$, we obtain the characteristic length of the charge-current generation to be $\lambda_\mathrm{Ti} = 4.6$~nm. This characteristic length is an order of magnitude smaller than that of the orbital-torque generation~\cite{hayashi2022observation,choi2021observation}, implying that the length scale associated with the orbital transport triggered by the orbital pumping is different from that triggered by the OHE. Similar arguments have been made for spin transport; a potential difference in the length scales of diffusive spin currents and intrinsically generated spin Hall currents has been suggested, but their relationship remains unclear~\cite{doi:10.1063/5.0024019}.

It is also worth noting the potential role of the inverse orbital Rashba-Edelstein effect in the charge-current generation induced by the orbital pumping. Recently, ultrafast dynamics of orbital transport have been studied using time-domain THz emission spectroscopy~\cite{seifert2023time}. In the experiment, the orbital transport is triggered by exciting a SiO$_2$/W/Ni film with femtosecond laser pulses. In the SiO$_2$/W/Ni film, the current-induced orbital torque, or the charge-to-orbital conversion, has been observed to be dominated by the OHE in the W layer~\cite{hayashi2022observation}. However, the THz emission dynamics in the SiO$_2$/W/Ni film indicates that the orbital-to-charge conversion is dominated by the inverse orbital Rashba-Edelstein effect at the SiO$_2$/W interface. This finding suggests that the direct and inverse orbital responses could be dominated by the different mechanisms. The relatively short characteristic length observed for the orbital pumping suggests the possibility that the inverse orbital Rashba-Edelstein effect contributes to the charge-current generation. Understanding the length scales associated with orbital transport remains a challenge for future studies.

\bigskip\noindent
\textbf{Conclusions and outlook}\\
In conclusion, we have presented experimental evidence for the orbital pumping, where the precession of magnetization results in the emission of an orbital current into an adjacent layer. 
We have shown that the orbital current generated by the orbital pumping in the Ni/Ti bilayer is converted into a charge current through the IOHE in the Ti layer without using spin-orbit coupling. 
Although a recent study on Y$_3$Fe$_5$O$_{12}$/Pt/CuO$_x$ structures suggested the involvement of the inverse orbital Rashba-Edelstein effect in the generation of a charge current, the underlying mechanism is the spin pumping into the Pt layer and subsequent spin-to-orbital conversion by spin-orbit coupling in the Pt layer~\cite{PhysRevApplied.19.014069}. This mechanism is distinct from the emission of orbital currents by the orbital pumping, observed in the present work.

The spin pumping is one of the most versatile and powerful mechanisms for spin-current generation, which enables spin injection into a wide variety of systems ranging from metals and semiconductors to magnetic insulators, organic materials, superconductors, and topological insulators~\cite{kajiwara,ando2011electrically,PhysRevLett.113.097202,watanabe2014polaron,jeon2018enhanced,PhysRevLett.113.196601}. This exceptional ability of the spin pumping has played an important role in the discovery of a variety of fundamental phenomena in spintronics. 
We anticipate that the orbital pumping---the orbital counterpart of the spin pumping---provides a powerful way to explore the physics of orbital currents and orbital dynamics. 
Our results provide a fundamental piece of information for deeper understanding of the angular momentum dynamics in solids at the nanoscale, which will stimulate further experimental and theoretical studies.

\clearpage

\clearpage

\bigskip\noindent
\textbf{Methods}\\
\textbf{Device fabrication.}
The FM(5 nm)/NM($t_{\rm{NM}}$) bilayers (FM =  Ni, Fe or Co, NM = Ti or Pt) were deposited on thermally oxidized Si substrates using radio frequency (RF) magnetron sputtering under a 6N-purity-Ar atmosphere at room temperature. The working pressure was 0.4 Pa. The thicknesses of the SiO$_2$ and Si layers are 100~nm and 625~$\mu$m, respectively. For the sputtering of the FM/Ti bilayers, the base pressure was better than $5\times 10^{-7}~\rm{Pa}$, which was achieved by reducing the residual hydrogen and oxygen contents through a sputtering process using a Ti target (at least 0.4~Pa, 5~min, 120~W). The FM/Pt bilayers were sputtered with a base pressure better than $1.0 \times 10^{-5}~\rm{Pa}$. To vary the Ti-layer thickness of the Ni/Ti bilayer on the same substrate, a wedged Ti film was fabricated by using a linear shutter. 
All samples were capped by 4-nm-thick-SiO$_2$ to prevent oxidation. For the pumping measurement, the films were fabricated into Hall cross structures with a width of $w=8~ \rm{\mu m}$ utilizing the conventional photolithography with negative resist followed by the Ar-ion milling and lift-off technique.
The device was placed under a coplanar waveguide (CPW) consisting of Ti(3 nm)/Au(400 nm) deposited by RF magnetron sputtering. A SiO$_2$(170~nm) insulating layer was deposited by RF magnetron sputtering between the device and the CPW.

\bigskip\noindent
\textbf{Microwave absorption and voltage measurements.}
For the FM/NM bilayers, we measured DC electromotive force $V_{\rm{DC}}$ using a nanovoltmeter (2182A, Keithley) with applying an RF current to the CPW and in-plane magnetic field $H$. The in-plane magnetic field $H$ was applied at an angle of $\theta$ from the direction of the applied RF current (see Fig.~\ref{fig1}b). The microwave absorption intensity $P$ was determined by measuring $S_{11}$ using a vector network analyzer (VNA) (N5222A, Keysight)~\cite{Iguchi_2012}, where $P=({\Delta|S_{11}|^2}/{|S_{11}^0|})P_{\mathrm{in}}$. Here, $|S_{11}^0|$ represents the reflection loss between the CPW and the port of the VNA without the FMR, and $\Delta |S_{11}|^2$ is the change ratio of the reflected microwave power at the FMR (see Supplementary Note 2).

\vspace{1cm}

\noindent\textbf{Data availability}\\
The data that support the findings of this study are available from the corresponding author upon reasonable request.

\clearpage

\textbf{References}\\

\clearpage

\bigskip\noindent
Correspondence and requests for materials should be addressed to K.A. (ando@appi.keio.ac.jp)\\

\bigskip\noindent
\textbf{Acknowledgements}\\
This work was supported by JSPS KAKENHI (Grant Number: 22H04964, 20H00337, 20H02593, 23K19037), Spintronics Research Network of Japan (Spin-RNJ), and MEXT Initiative to Establish Next-generation Novel Integrated Circuits Centers (X-NICS) (Grant Number: JPJ011438). H.H. is supported by JSPS Grant-in-Aid for Research Fellowship for Young Scientists (DC1) (Grant Number 20J20663). D.G. and Y.M. acknowledge Deutsche Forschungsgemeinschaft (DFG, German Research Foundation) - TRR 173/2 - 268565370 - Spin+X (Project A11), and TRR 288 - 422213477 (Project B06), for funding.

\bigskip\noindent
\textbf{Competing interest}\\
The authors declare no competing interests.

\bigskip\noindent
\textbf{Author contributions}\\
H.H. fabricated devices, collected and analyzed the data, and performed the materials characterization. H.H. and K.A. designed the experiments. H.H., D.G., Y.M., and K.A. developed the explanation. H.H. and K.A. wrote the manuscript with the help from D.G and Y.M. All authors discussed results and reviewed the manuscript. K.A. supervised the study.

\color{black}

\end{document}